\documentclass[aps,pre,amsmath,amssymb,lengthcheck,showpacs,superscriptaddress]{revtex4-1}
\usepackage{graphicx}
\begin{document}
\title{Acoustic excitations and elastic heterogeneities in disordered solids}
\author{Hideyuki Mizuno}
\affiliation{Laboratory for Interdisciplinary Physics, UMR 5588, Universit\'e Grenoble 1 and CNRS, 38402 Saint Martin d'H\`eres, France}
\author{Stefano Mossa}
\affiliation{CEA, INAC, SPrAM, UMR 5819 (UJF, CNRS, CEA), 17 rue des Martyrs, 38054 Grenoble Cedex 9, France}
\author{Jean-Louis Barrat}
\affiliation{Laboratory for Interdisciplinary Physics, UMR 5588, Universit\'e Grenoble 1 and CNRS, 38402 Saint Martin d'H\`eres, France}
\affiliation{Institut Laue-Langevin - 6 rue Jules Horowitz, BP 156, 38042 Grenoble, France}
\date{\today}
\begin{abstract}
In the recent years, much attention has been devoted to the inhomogeneous nature of the mechanical response at the nano-scale in disordered  solids. Clearly, the elastic heterogeneities that have been characterized in this context are expected to strongly impact the nature of the sound waves which, in contrast to the case of perfect crystals, cannot be  completely rationalized in terms of phonons. Building on previous work on a toy model showing an amorphisation transition [Mizuno H, Mossa S, Barrat JL (2013) EPL {\bf 104}:56001], we investigate the relationship between sound waves and elastic heterogeneities in a unified framework, by continuously interpolating from the perfect crystal, through increasingly  defective phases, to fully developed glasses. We provide strong evidence of a direct correlation between sound waves features and the extent of the heterogeneous mechanical response at the nano-scale. 
\end{abstract}
\maketitle
\section{Introduction}
In crystals, molecules thermally oscillate around the periodic lattice sites and vibrational excitations are well understood in terms of quantized plane waves, the phonons~\cite{kettel}. The vibrational density of states (vDOS) in the low frequency regime is well described by the Debye model, where the vibrational modes are the acoustic phonons. In contrast, disordered solids, including structural glasses and disordered crystals, exhibit specific  vibrational properties compared to the corresponding pure crystalline phases. In particular, the origin of the vDOS modes in excess over the Debye prediction around $\omega\sim 1\ \text{THz}$, the so-called Boson Peak (BP), is still debated (see, among many others,~\cite{buchenau_1984,lowtem}). At the BP frequency, $\Omega^\text{BP}$, localized modes have also been observed~\cite{mazzacurati_1996}. Acoustic plane waves, which are {\em exact} normal modes in crystals, can still propagate in disordered solids. Indeed, at low frequencies, $\Omega$, and long wavelengths, $\Lambda$, acoustic sound waves do not interact with disorder and can propagate conforming to the expected macroscopic limit. However, as $\Omega$ is increased beyond the Ioffe-Regel (IR) limit, $\Omega^\text{IR}$, acoustic excitations interact with the disorder and are strongly dumped~\cite{monaco_2009,monaco2_2009,marruzzo2013}. Interestingly, this strong scattering regime occurs around the BP position,  $\Omega^\text{IR} \sim \Omega^\text{BP}$~\cite{ruffle_2006,shintani_2008}. The exact origin of strong scattering and its connection to the BP remains elusive.

A possible rationalization of the above issues is based on the existence of {\em elastic heterogeneities}~\cite{Duval_1998}, which can originate from structural disorder, as in structural glasses~\cite{buchenau_1984}, or disordered inter-particle potentials, even in lattice structures such as disordered colloidal crystals~\cite{Kaya_2010}. Within the framework of jamming approaches and using effective medium theories, elastic heterogeneities are related   to the proximity of local elastic instabilities \cite{degiuli2014}. Recent simulation work~\cite{yoshimoto_2004,tsamados_2009,Mizuno_2013} has clearly demonstrated their existence in disordered solids.  This is at variance with the case of simple crystals, which are characterized by a fully affine response and homogeneous moduli distributions~\cite{Wagner_2011}. More specifically, in the large length scale limit, macroscopic moduli are observed. In contrast, as the length-scale is reduced, moduli heterogeneities are detected, at a typical length scale $\xi\simeq 10-15\sigma$~\cite{tsamados_2009}, where $\sigma$ is the typical atomic diameter. Breakdown of both continuum mechanics~\cite{wittmer_2002} and Debye approximation~\cite{monaco_2009,monaco2_2009} has been demonstrated at the same \textit{mesoscopic} length-scale $\xi$, where they are still valid for crystals. Remarkably, the wave frequency corresponding to the wavelength $\Lambda \sim \xi$ is very close to $\Omega^\text{IR} \sim \Omega^\text{BP}$~\cite{leonforte_2005}.  Altogether these results indicate that a close connection must exist between elastic heterogeneities and acoustic excitations. In this Report we precisely address this point.

In Ref.~\cite{Mizuno2_2013} we considered a numerical model featuring an amorphisation transition~\cite{bocquet_1992}. We showed how to systematically deform the local moduli distributions, evaluated by coarse-graining the system in small domains of linear length-scale $w$. We characterized the degree of elastic heterogeneity in terms of standard deviation of those distributions, and studied the effect on normal modes (eigenvalues of the Hessian matrix) and thermal conductivity. Building on that work, we are now in the position to investigate the relation between elastic heterogeneities and acoustic excitations, unifying in a single framework ordered and disordered solid states. By interpolating in a controlled way from perfect crystals, through increasingly  defective phases, to fully developed amorphous structures, we: {\em i)} calculate the dynamical structure factors, extracting the relevant spectroscopic parameters; {\em ii)} characterize the wave vector dependence of sound velocity and broadening of the acoustic excitations and clarify their nature in terms of the Ioffe-Regel limit; {\em iii)} provide, for the first time, direct evidence of the correlation of the excitations life-times and $\Omega^\text{IR}$ with the magnitude of the elastic heterogeneities.
\section{Results and Discussion}
We study by Molecular Dynamics simulation in the $NVT$ ensemble, at constant temperature $T=0.01$ and number density $\hat{\rho}=N/V=1.015$ ($V$ being the system volume), a $50:50$ mixture, composed by $N$ atoms with different diameters, $\sigma_{1}$ and $\sigma_{2}$, and same mass, $m=1$. We consider two different system sizes $N=$~108,000 and 256,000, to improve statistics and wave vector range, and confirm that results are not affected by finite-size effects. Particles interact via a soft-sphere potential, $v_{\alpha \beta}=\epsilon (\sigma_{\alpha \beta}/r)^{12}$, with $\sigma_{\alpha \beta}=(\sigma_{\alpha}+\sigma_{\beta})/2$ and $\alpha, \beta \in 1, 2$. The potential is cut off and shifted at $r = 2.5 \sigma_{\alpha \beta}$. In a one component approximation, we define an ``effective" diameter $\sigma_{\text{eff}}^3 = \sum_{\alpha,\beta=1,2} \sigma_{\alpha \beta}^3$/4~\cite{bernu_1987}. Starting from a perfect face-centered cubic crystal, defects are added in the form of size disorder, by simultaneously decreasing $\sigma_1$ below the initial value $\sigma_1=1$ and increasing $\sigma_2$, keeping a constant  $\sigma_\text{eff}\equiv 1$~\cite{bocquet_1992}. The size ratio, $\lambda = \sigma_1/\sigma_2 \le 1$, quantifies the size disorder and is our control parameter. $\lambda=1$ corresponds to the perfect crystal case, while  for $\lambda=0.7$ a completely developed amorphous structure is observed. An amorphisation transition occurs at $\lambda = \lambda^\ast \simeq 0.81$ \cite{Mizuno2_2013,bocquet_1992}. Additional details can be found in Ref.~\cite{Mizuno2_2013}. Simulations have been realized by using the large-scale, massively parallel molecular dynamics computer simulation code LAMMPS~\cite{plimpton95}.
%
%
\begin{figure}[tb]
\centering
\includegraphics[width=0.49\textwidth]{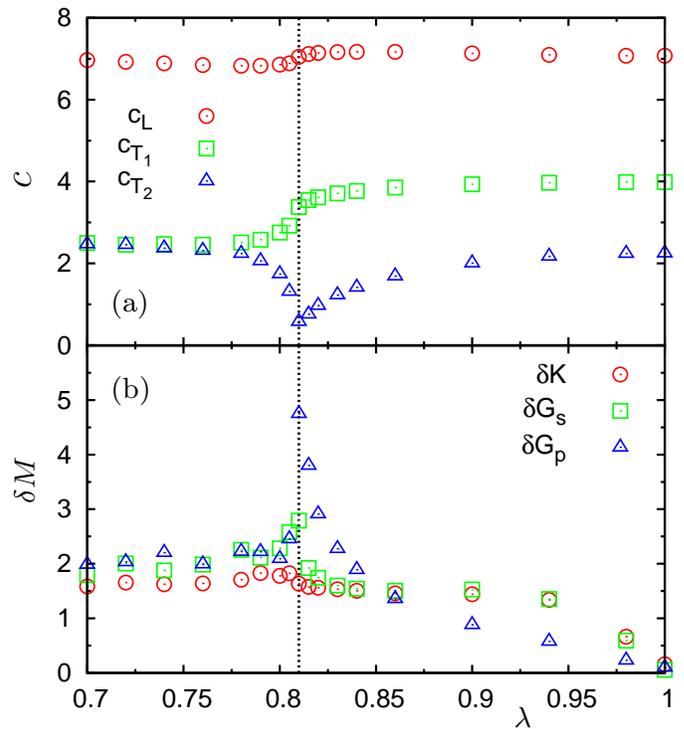}
\caption{
Macroscopic limit of sound velocities and width of the distributions of local elastic moduli. {\em a)} $\lambda$-dependence  of the longitudinal ($L$) and transverse ($T_1$, $T_2$) macroscopic sound velocities in the $(110)$-direction. These data have been calculated from the effective elastic moduli $K+G_p/3+G_s$, $G_s$, and $G_p$, respectively. Here $K$, $G_p$, and $G_s$ are the bulk, pure shear and simple shear moduli. The vertical dashed line indicates the transition point $\lambda=\lambda^\ast \simeq 0.81$. {\em b)}  $\lambda$-dependence of the elastic heterogeneities, $\delta M$ associated to $K$, $G_s$, and $G_p$. These data are the standard deviations of the distribution of the local elastic moduli, for a coarse-graining length scale $w=3.16$~\cite{Mizuno2_2013}. 
} 
\label{soundspeed}
\end{figure}
%
%
\begin{figure}[tb]
\centering
\includegraphics[width=0.49\textwidth]{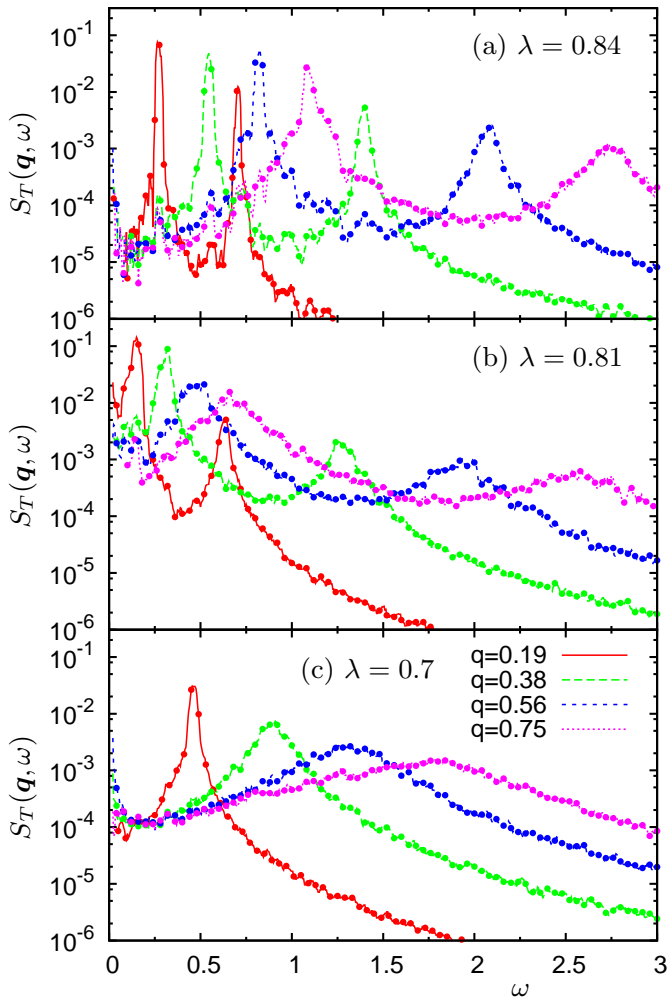}
\caption{
Transverse dynamic structure factors $S_{T}(\vec{q},\omega)$, at the indicated values of the wave vector $\vec{q}$  in the $(110)$-direction, calculated from Eq.~(\ref{eq:sqw}). Three values of $\lambda$ are shown, in a defective crystal state {\em a)}, at the amorphisation transition {\em b)}, and in the fully developed glassy phase {\em c)}. Two Brillouin peaks, corresponding to the $T_1$ and  $T_2$ branches, are visible for $\lambda=0.84$ and $0.81$. In the glassy phase ($\lambda=$0.7) only one degenerate excitation survives. 
} 
\label{dynamic}
\end{figure}

We first focus on the acoustic sound velocities in the macroscopic limit. In crystals, sound propagation depends on the direction of the wave-vector, $\vec{q}$~\cite{kettel}. This is at variance with the isotropic amorphous phases, where only the wave vector modulus is relevant. In the macroscopic limit, the $q$-independent sound velocity is $c=\sqrt{M_\text{eff}/\rho}$, where $\rho$ is the mass density, and $M_\text{eff}$ is an effective macroscopic modulus which depends on both the direction of propagation and the longitudinal or transverse character of the excitation. In what follows we will consider the $(110)$-direction, with $M_\text{eff}=K+G_p/3+G_s$, $G_s$, and $G_p$, for the longitudinal ($L$) and the two transverse ($T_1$ and $T_2$) branches, respectively. Here $K$, $G_p$, and $G_s$ are the bulk, pure shear and simple shear moduli~\cite{Mizuno_2013}. Additional results for the $(100)$ and $(111)$ directions are reported in the SI. In Fig.~\ref{soundspeed}~a) we show the $\lambda$-dependence of $c$ for the three branches. For $\lambda>\lambda^\ast$, $c_{T_1}>c_{T_2}$, and both slowly follow the decrease of $\lambda$. At $\lambda^\ast$, $c_{T_2}\simeq 0$, which can be associated with an elastic instability controlled by $G_p$~\cite{Mizuno2_2013}. For $\lambda<\lambda^\ast$, $c_{T_1}$ decreases while $c_{T_2}$ increases and both reach the same values in the fully developed amorphous state, as expected. Note that glass and pure crystal show very similar $c_{T_2}$ in the macroscopic limit. Finally, in the entire $\lambda$-range, the overall variation of $c_{L}$ is very mild.

In Fig.~\ref{soundspeed}~b) we also display  the  $\lambda$-dependence of the standard deviation, $\delta M$, calculated from the probability distributions of the local moduli, $M=K$, $G_s$, and $G_p$, respectively. These can be evaluated by coarse-graining the system in little cubic domains, of linear size $w=3.16$ in this case~\cite{Mizuno2_2013}. Starting from a spatially homogeneous distribution at $\lambda=1$, $\delta G_p$ undergoes very important modifications, strongly increasing by decreasing $\lambda$, reaching a maximum at $\lambda^\ast$, and abruptly decreasing to a stable low value on the amorphous side. $\delta G_s$ follows a qualitatively similar behaviour, while quantitatively less important, and the expected degeneracy is recovered in the amorphous phases. Finally, longitudinal data also undergo variations similar to those of $\delta G_s$ for $\lambda\ge 0.82$, eventually staying almost unchanged across the transition.

Moving from the macroscopic limit, we now investigate the wave-vector dependence of the dynamic structure factors,
\begin{equation}
S_a(\vec{q},\omega) = \frac{1}{2 \pi N} \left( \frac{q}{\omega} \right)^2 \int dt \left< \vec{j_a}(\vec{q},t) \vec{j^\ast_a}(\vec{q},t) \right> e^{i \omega t},
\label{eq:sqw}
\end{equation}
where $a=L, T$, and $\vec{j_L}(\vec{q},t)$ and $\vec{j_T}(\vec{q},t)$ are the longitudinal and transverse momentum currents, respectively~\cite{monaco2_2009,shintani_2008}. It is by now consensual that transverse modes play the most important role in determining anomalies in vibrational properties~\cite{shintani_2008}. More specifically, the transverse branch with the lowest elastic modulus has been demonstrated to be the one which correlates most to the low-frequency vibrational states~\cite{Mizuno2_2013}. In what follows we therefore focus on the $T_2$ excitations. Additional data for the $T_1$ and $L$ modes are included in the SI. In Fig.~\ref{dynamic} we plot $S_{T}(\vec{q},\omega)$, at the indicated values of $\lambda$ and $q$. For $\lambda=0.84$ and $0.81$, where the two transverse sound velocities are well separated (Fig.~\ref{soundspeed}~a)), $S_{T}(\vec{q},\omega)$ features two Brillouin peaks corresponding to $T_1$ (high-$\omega$) and $T_2$ (low-$\omega$) excitations, respectively. In contrast, a single Brillouin peak is visible, as expected,  in the amorphous phase at $\lambda =0.7$, where $c_{T_1}\simeq c_{T_2}$.
%
%
\begin{figure}[tb]
\centering
\includegraphics[width=0.49\textwidth]{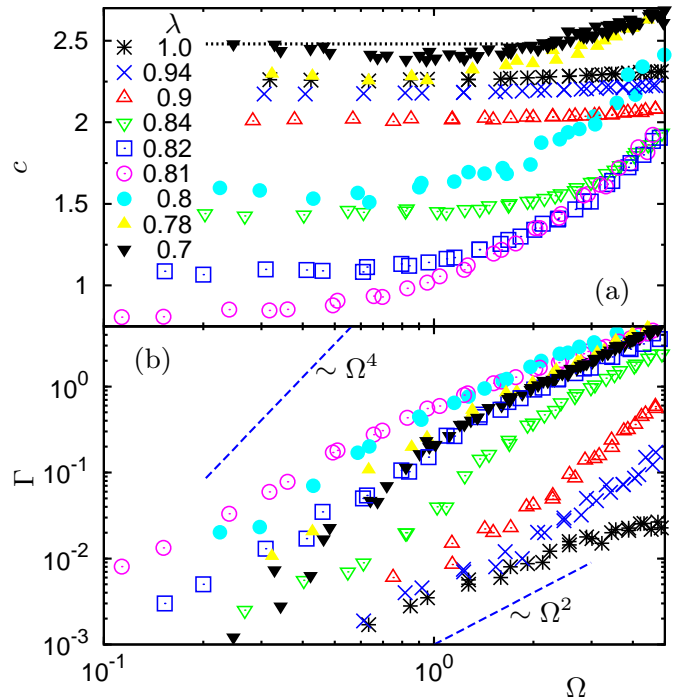}
\caption{
Spectroscopic parameters calculated from the dynamic structure factors $S_{T}(\vec{q},\omega )$.
{\em a)} Transverse phase velocity $c(\Omega)= \Omega (q)/q$, and {\em b)} broadening, $\Gamma (\Omega)$, for the $T_2$ excitations in the direction $[110]$, at the indicated values of $\lambda$. These data have been obtained by fitting the calculated $S_{T}(\vec{q},\omega )$ to the damped harmonic oscillator line shape of Eq.~(\ref{eq:dho}). The horizontal dashed line in {\em a)} corresponds to the macroscopic limit of the sound velocity at $\lambda=0.7$. The dashed lines $\propto \Omega^2$ and $\propto \Omega^4$ in {\em b)} are also guides for the eye, to emphasize the extremely complex frequency-dependence of $\Gamma$ at different values of $\lambda$. A comprehensive discussion of these data is included in the main text.
} 
\label{gamma}
\end{figure}
%
%
\begin{figure}[tb]
\centering
\includegraphics[width=0.49\textwidth]{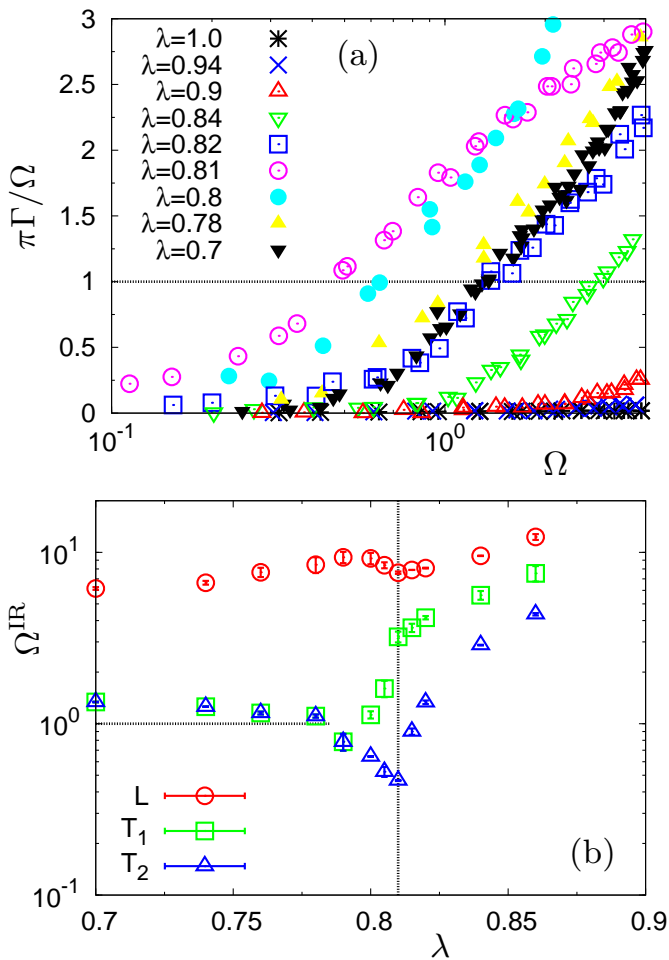}
\caption{
Characterization of the Ioffe Regel limit. {\em a)} Ratio $\pi \Gamma / \Omega$ at the indicated values of $\lambda$, calculated from the data of Fig.~\ref{gamma}, for the $T_2$ excitations in the $[110]$ direction. The frequency corresponding to the intersection of each data set with the horizontal solid line at the value one defines the  Ioffe-Regel limit, $\Omega^\text{IR}$.  {\em b)} $\lambda$-dependence of $\Omega^\text{IR}$ extracted from the above data, corresponding to the $T_2$ branch. We also show $\Omega^\text{IR}$ for the longitudinal ($L$) and higher transverse ($T_1$) branches, respectively. The vertical line indicates the transition point $\lambda^\ast \simeq 0.81$. The horizontal line corresponds to the boson peak position, $\Omega^\text{BP}\simeq 1$, for the amorphous phases at $\lambda \le 0.78$.
} 
\label{irlimit}
\end{figure}

Propagation frequency, $\Omega_a(\vec{q})$, and line broadening, $\Gamma_a(\vec{q})$, of the sound excitations can be extracted from these data by fitting the spectral region around the Brillouin peaks to a damped harmonic oscillator model~\cite{monaco2_2009,shintani_2008},
\begin{equation}
S_a(\vec{q},\omega) \sim \frac{\Gamma_a(\vec{q})\Omega^2_a(\vec{q})}{ (\omega^2-\Omega^2_a(\vec{q}))^2 + \omega^2\Gamma^2_a(\vec{q})}.
\label{eq:dho}
\end{equation}
In Figs.~\ref{gamma}(a) and (b) we show the sound velocity, $c=c_{T_2}= \Omega_{T_2}(\vec{q})/q$, and broadening, $\Gamma=\Gamma_{T_2}$, at the indicated values of the disorder parameter $\lambda$. For the sake of clarity, we consider first the isotropic amorphous case, $\lambda=0.7$. As expected, for vanishing $\Omega$, $c$ corresponds to the macroscopic value of Fig.~\ref{soundspeed}(a) (horizontal dashed line), calculated directly from the value of $G_p$ at the same $\lambda$-value. Next, $c(\Omega)$ decreases ({\em softening}), reaches a minimum, and eventually undergoes positive dispersion at higher frequencies. In the same region where $c$ shows a minimum, a crossover from $\sim \Omega^2$ at high frequency 
(which can be described by a two-mode Maxwell constitutive model~\cite{mizuno_yamamoto_2013}), to a Rayleigh-like $\sim \Omega^\alpha$ with $\alpha$ close to $4$ at intermediate frequency is evident for $\Gamma$ around $\Omega\simeq 1$, which corresponds to $\Omega^\text{BP}$ in this case~\cite{Mizuno2_2013}. Both these features are consistent with previous findings for the Lennard-Jones glass~\cite{monaco2_2009,marruzzo2013}.

As $\lambda$ increases, the  sound velocity at a given frequency first decreases, goes through a {\em minimum} at $\lambda^\ast \simeq 0.81$, and eventually increases steadily. We note that the maximum ratio $\simeq 3.5$ between the maximum and minimum value (as a function of frequency)  is reached at $\lambda^\ast$, while $c$ is essentially frequency independent at $\lambda\ge 0.9$, where the Debye picture still holds. Therefore, $c$ mirrors at all frequencies the non-monotonic behaviour of the macroscopic limit of Fig.~\ref{soundspeed}(a). Sound broadening follows a quite different pattern. As $\lambda$ increases from $0.7$,  $\Gamma$ is enhanced and reaches a {\em maximum} at $\lambda^\ast$. Next, it is strongly suppressed for $\lambda > \lambda^\ast$, converging to a very low value at $\lambda=1$, of pure anharmonic origin. We remark that in this case, the ratio between the maximum and minimum values reached, covers almost two decades at $\Omega \simeq 1$. We will see below that this finding can be rationalized in terms of a strong correlation with the magnitude of the elastic heterogeneity associated with the appropriate modulus (Fig.~\ref{soundspeed}(b)). 

Next, we focus on the Ioffe-Regel limit, $\Omega^\text{IR}$, for all investigated $\lambda$'s. In Fig.~\ref{irlimit}~(a) we propose a different representation of the data points of Fig.~\ref{gamma}~(a) and~(b), as the ratio $\pi \Gamma (\Omega)/ \Omega$. At the IR limit, $\pi \Gamma (\Omega^\text{IR})/ \Omega^\text{IR}=1$, {\em i.~e.}, the decay time of the excitations equals half of the corresponding vibrational period. $\Omega^\text{IR}$ provides an upper bound for the validity of acoustic-like descriptions of the vibrational excitations. The $\lambda$-dependence of $\Omega^\text{IR}$ presents again an interesting non-monotonous pattern, which we make quantitative in Fig.~\ref{irlimit}~(b). Here we plot the $\Omega^\text{IR}$ extracted from the above data, together with our results for the other two branches, ${T_1}$ and $L$. Starting from the pure crystal, where $\Omega^\text{IR}$ is expected to be comparable to the highest frequency comprised in the vDOS, the IR limit decreases steadily with $\lambda$ in all cases, reaches a minimum at $\lambda^\ast$ for $T_2$ and $L$ (in the $T_1$ case $\Omega^\text{IR}$ continuously decreases through the transition) and levels off  to a constant value on the amorphous side. For the two transverse branches, this value corresponds to the $\Omega^\text{BP}$ position, while for the longitudinal mode $\Omega^\text{IR} \gg \Omega^\text{BP}$, as already shown in Refs.~\cite{monaco2_2009,shintani_2008}. Note that a recent study~\cite{Duval_2013} reported that the nature of the BP depends on the Poisson ratio, $\nu$: for fragile glasses with relatively high $\nu>0.25$, $\Omega^\text{BP}\sim\Omega^\text{IR}_T$~\cite{monaco2_2009,shintani_2008}, whereas for strong glasses with lower $\nu<0.2$, $\Omega^\text{BP}\sim\Omega^\text{IR}_L$~\cite{ruffle_2006}. We have checked that our soft-sphere model is a fragile system, with $\nu \simeq 0.43$ for $\lambda \le 0.78$, which is consistent with these findings. Unfortunately, for $\lambda>\lambda^\ast$ we were not able to determine reliably the value of $\Omega^\text{BP}$. However, in Ref.~\cite{Mizuno2_2013} (Figs.~3(b)) we showed that $\Omega^\text{BP}$ shifts to lower frequencies as $\lambda$ tends to $\lambda^\ast$ from above, and increases back to higher frequencies below the transition. This behaviour clearly mirrors the pattern followed by $\Omega^\text{IR}$ of ${T_2}$ excitations in Fig.~\ref{irlimit}(b), implying the lowest-frequency $T_2$ excitations are most related to the Boson peak. Therefore, we can conjecture that $\Omega^\text{IR} \sim \Omega^\text{BP}$ above the transition also.
%
%
\begin{figure}[tb]
\centering
\includegraphics[width=0.49\textwidth]{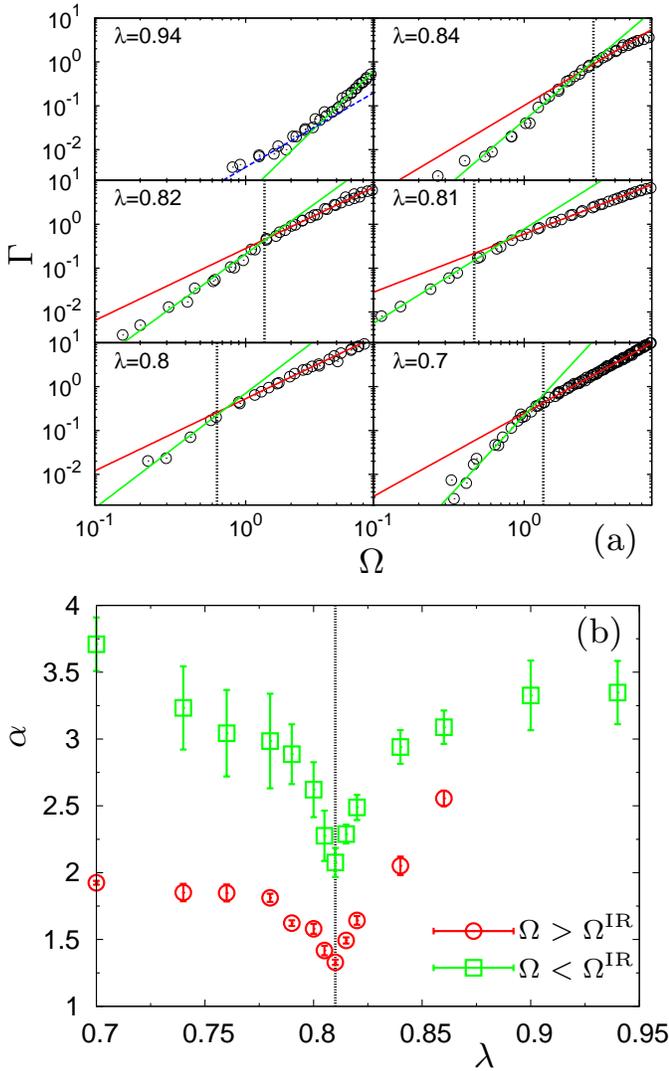}
\caption{
Frequency dependence of broadening for the $T_2$ excitations, at the investigated values of $\lambda$ . {\em a)} $\Omega$-dependence of the broadening $\Gamma$ at the indicated values of $\lambda$. Red and green solid lines are the best power-law fits of the form $\simeq \Omega^\alpha$ to the data, in the high ($\Omega > \Omega^\text{IR}$) and intermediate ($\Omega < \Omega^\text{IR}$) frequency ranges respectively. $\Omega^\text{IR}$ is indicated by the vertical dashed lines in all cases except $\lambda=0.94$, where $\Omega^\text{IR}$ is comparable to the highest frequency comprised in the vDOS. For $\lambda=0.94$, the low-frequency cross-over to the anharmonic $\simeq \Omega^2$ behaviour is indicated by the blue dashed line. {\em b)} Values of the exponents $\alpha$ extracted from the best fit to the data of the {\em a)} panel, for the high (circles) and intermediate (squares) frequency-ranges, respectively. A detailed discussion of these data is included in the text.
}
\label{exponent}
\end{figure}
%
%
\begin{figure}[tb]
\centering
\includegraphics[width=0.49\textwidth]{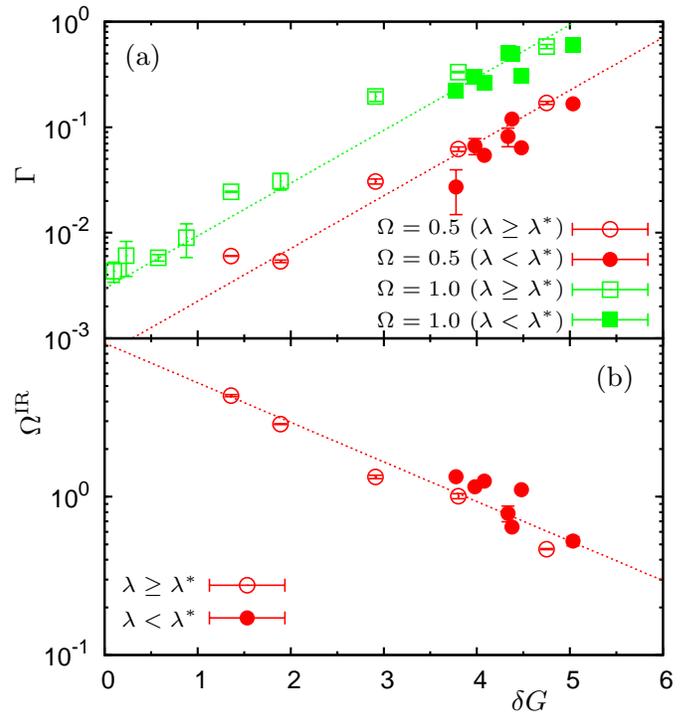}
\caption{
Direct correlation of features of acoustic-like excitations with local elastic heterogeneities. Here we show the dependence on the extent of the elastic heterogeneities of {\em a)} broadening, $\Gamma$, and {\em b)} Ioffe-Regel frequency, $\Omega^\text{IR}$, for the $T_2$ acoustic excitations. Data are plotted versus the standard deviation $\delta G$ of the distribution of the relevant local elastic modulus, calculated for a coarse-graining length scale $w=3.16$ (see Fig.~\ref{soundspeed}(b)). $\delta G\simeq \delta G_p$ for $\lambda\ge \lambda^\ast$ (open symbols),  $\delta G\simeq \delta G_p+\delta G_s$ for $\lambda < \lambda^\ast$, in the amorphous phases (closed symbols). A discussion of this point is included in the main text. In {\em a)} we plot values of $\Gamma$ corresponding to two different fixed frequencies, $\Omega\simeq 0.5$ and $\Omega\simeq 1$. Dashed lines are guides for the eye.  
}
\label{gamma-vs-dg}
\end{figure}

The data shown in Fig.~\ref{gamma}(b) acquire even more interest in light of the above discussion of the Ioffe-Regel limit. Indeed, the shape of the $\Gamma(\Omega)$ functions significantly changes with $\lambda$, and follows quite complex patterns. These show, at frequencies close to $\Omega^\text{IR}$, clear cross-overs between regimes with different effective exponents, $\alpha$, at high and intermediate frequencies~\cite{monaco2_2009}. The expected low-frequency cross-over to the $\simeq \Omega^2$ behaviour due to anharmonicity can be recognized for $\lambda \ge 0.94$ (also see the blue dashed line in Fig.~\ref{exponent}~a)), while it cannot be observed for lower values of $\lambda$ at the considered temperature. In Fig.~\ref{exponent}~a) the red and green solid lines are the best power-law fits to the data, in the high ($\Omega > \Omega^\text{IR}$) and intermediate ($\Omega < \Omega^\text{IR}$) frequency ranges, respectively. The positions of $\Omega^\text{IR}$ are indicated by the vertical dashed lines, for the cases $\lambda<0.94$. For $\lambda=0.94$, $\Omega^\text{IR}$ is already close to the highest frequency comprised in the vDOS. The obtained values of the exponents $\alpha$ in the two regimes are shown in Fig.~\ref{exponent}~b). For $\Omega < \Omega^\text{IR}$, and for the deeply amorphous state $\lambda=0.7$, $\alpha\simeq 3.7$, compatible with the expected Rayleigh scattering exponent $\alpha=4$. By increasing $\lambda$, $\alpha$ first decreases steadily by reaching the value $2$ at $\lambda^\ast$, next increases up to a value $\simeq 3.5$ at $\lambda=0.94$. For $\Omega > \Omega^\text{IR}$, we recover the expected value $\alpha=2$ in the amorphous phase \cite{mizuno_yamamoto_2013}, which decreases quite abruptly, reaching a value $\simeq 1.5$ at $\lambda^\ast$. Subsequently, $\alpha$ increases to a value close to $2.5$ at $\lambda=0.86$.  These data, in particular those corresponding to the high frequency branch, provide information similar to that of Ref.~\cite{angelani_2000}. There, it was shown by a quite involved analysis that {\em frustration} seems to control the value of $\alpha$. More precisely, $\alpha \simeq 4$ when frustration is absent, and decreases even below the value $2$ by increasing frustration. This is consistent with our findings which, however, provide a broader picture, including predictions on the low-frequency values. Note that, at variance with Ref.~\cite{angelani_2000}, here we can refer to topologically ordered and disordered systems described by the same family of Hamiltonians.

We now stress an interesting feature emerging from our results. The total variation of the IR limit for the $T_2$ branch on approaching $\lambda^\ast$ from above is very large (an order of magnitude), and $\Omega^\text{IR}$ apparently is (anti-)correlated with the elastic heterogeneities of Fig.~\ref{soundspeed}(b). Above $\lambda^\ast$, we noticed that the sound broadening also has a quite large overall variation and seems to follow the evolution of the elastic heterogeneities. We make quantitative these correlations in Fig.~\ref{gamma-vs-dg}, which is the most relevant result of this work. In Fig.~\ref{gamma-vs-dg}~(a) we plot $\Gamma$ on both sides of the transition, at the low frequencies $\Omega\simeq 0.5$ and $1$, and as a function of the extent of the elastic  heterogeneities at the corresponding $\lambda$ (Fig.~\ref{soundspeed}~(b)). While in the non-degenerate cases $\lambda>\lambda^\ast$ (open symbols), the appropriate data to consider are $\delta G\simeq \delta G_p$, in the amorphous cases (filled symbols), where the transverse moduli are degenerate, we assume additivity of the disorder sources and use $\delta G\simeq\delta G_p+\delta G_s$. Remarkably, the data follow an exponential behavior $\Gamma\propto \exp (\delta G/g_\Gamma)$ for both frequencies. Similarly, we find $\Omega^\text{IR}\propto \exp (-\delta G/g_{\Omega^\text{IR}})$ (Fig.~\ref{gamma-vs-dg}~(b)), for both lattice and amorphous cases. Note that no adjustable parameters are involved in these plots. Although we cannot be more precise on the origin of the particular functional form, we can conservatively assert that these data are the first strong evidence of a {\em direct} correlation of quantities related to the intrinsic nature of acoustic-like excitations in ordered/defective/amorphous phases with local mechanical properties at the nano-scale, {\em i.~e.}, heterogeneity of the elastic moduli. 
\section{Conclusions}
In summary, in this work we have investigated sound waves propagation in a numerical model featuring an amorphisation transition. By controlling the extent of a well-designed form of size disorder, we have been able to consider a panoply of different solid states of matter, ranging from the perfect crystal and increasingly defective lattice structures, to completely amorphous phases. This approach can be seen as a numerical analogue of experiments that  compare  scattering experiments on glasses and the corresponding (poly-)crystalline polymorphs~\cite{chumakov2014role,baldi2013emergence}. By calculating the appropriate dynamical structure factors, we have fully characterized transverse and longitudinal vibrational excitations in terms of sound velocities and broadening, also providing a very detailed analysis of the complex frequency dependence of the latter. Both the life-time and the Ioffe-Regel limit of the sound-like excitations have been demonstrated to directly correlate with the width of the distributions of local elastic moduli, both in the cases of lattice systems with defects and isotropic amorphous structures. These results provide the first direct evidence that elastic heterogeneities
crucially influence the most puzzling features in acoustic-like excitations in disordered systems, including strong scattering and Boson peak.
\begin{acknowledgments}
This work was supported by the Nanosciences Foundation of Grenoble. J.-L.~B is supported by the Institut Universitaire de France. Most of the computations presented in this paper were performed using the Froggy platform of the CIMENT infrastructure (https://ciment.ujf-grenoble.fr), which is supported by the Rh\^{o}ne-Alpes region (GRANT CPER07\_13 CIRA) and the Equip@Meso project (reference ANR-10-EQPX-29-01) of the programme Investissements d'Avenir supervised by the Agence Nationale pour la Recherche.
\end{acknowledgments}
\bibliographystyle{apsrev4-1}
\widetext
\section*{Supporting Information}
In what follows we report additional data, referring to the $(100)$ and $(111)$ directions in the wave vector space, for both longitudinal and transverse dynamical structure factors. These data are shown for completeness, the main article containing all results needed to support our conclusions.
\vspace{1cm}

\begin{figure*}[h]
\centering
\includegraphics[width=0.45\textwidth]{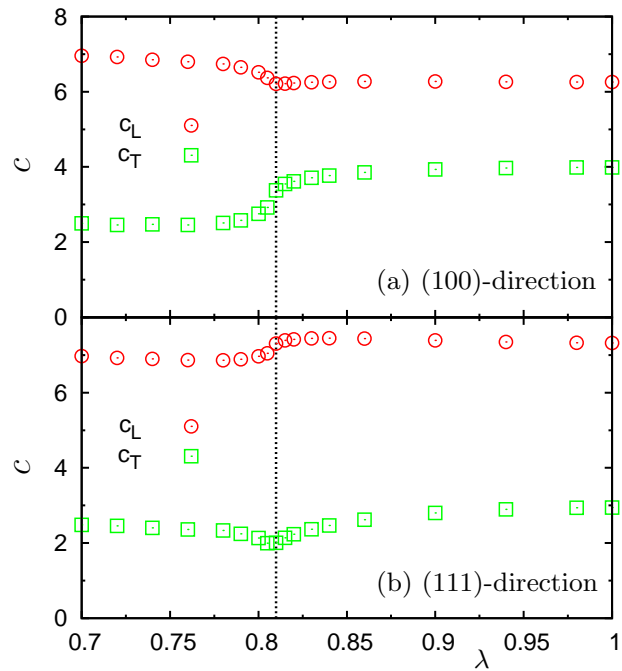}
\caption{
$\lambda$-dependence of the longitudinal ($L$) and transverse ($T$) macroscopic sound velocities, {\em a)} in the $(100)$-direction and
 {\em b)} in the $(111)$-direction. These data have been calculated from the effective elastic moduli, $K+4G_p/3$, $G_s$ in the $(100)$
-direction, and $K+4G_s/3$, $(2G_p+G_s)/3$ in the $(111)$-direction. The vertical dashed line indicates the transition point $\lambda=\
lambda^\ast \simeq 0.81$. Note that in the deep isotropic amorphous state $\lambda=0.7$, $c_L$ and $c_T$ assume the same values in the 
three directions, $(100)$, $(110)$, and $(111)$.
} 
\label{macro-velocity}
\end{figure*}
\begin{figure*}[]
\centering
\includegraphics[width=0.45\textwidth]{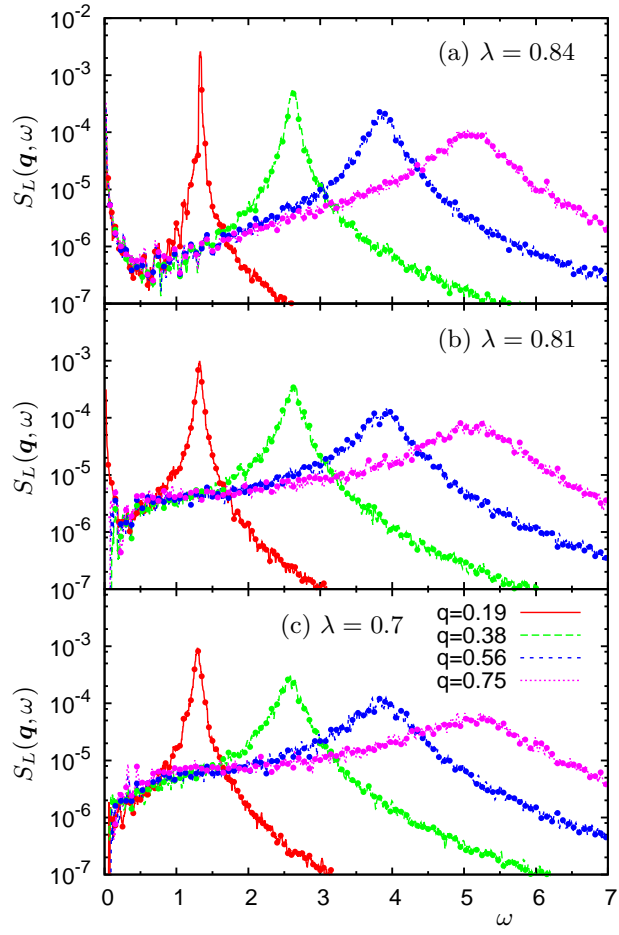}
\caption{
Longitudinal dynamic structure factor, $S_{L}(\vec{q},\omega)$, at the indicated values of the wave vector $\vec{q}$  in the $(110)$-di
rection, calculated from Eq.~(1) of the main text. Three values of $\lambda$ are shown, in a defective crystal state {\em a)}, at the a
morphisation transition {\em b)}, and in the fully developed glassy phase {\em c)}. Contrary to the case of the transverse dynamic stru
cture factors $S_{T}(\vec{q},\omega)$ shown in Fig.~2 of the main text, $S_{L}(\vec{q},\omega)$ features a single Brillouin peak even i
n the lattice cases.
} 
\label{dynamic}
\end{figure*}
\begin{figure*}[]
\centering
\includegraphics[width=0.45\textwidth]{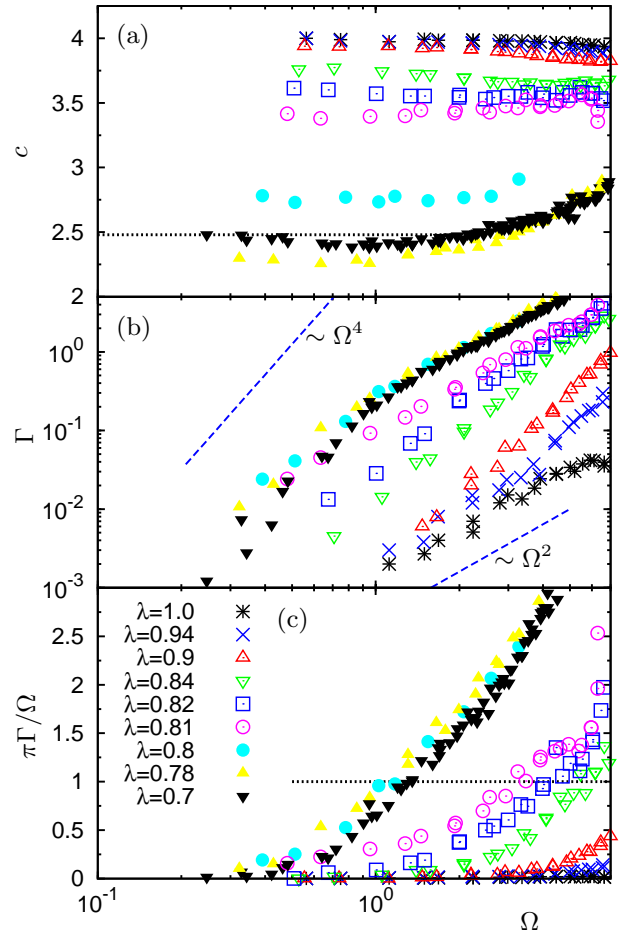}
\caption{
Spectroscopic parameters calculated from the transverse dynamic structure factors $S_{T}(\vec{q},\omega )$.
{\em a)} Phase velocity $c(\Omega)= \Omega(q)/q$, and {\em b)} broadening $\Gamma (\Omega)$, for the $T_1$ excitations in the direction
 $[110]$, at the indicated values of $\lambda$. The horizontal dashed line in {\em a)} corresponds to the macroscopic limit of the soun
d velocity at $\lambda=0.7$. The dashed lines $\propto \Omega^2$ and $\propto \Omega^4$ in {\em b)} are guides for the eye. {\em c)} Th
e ratio $\pi \Gamma(\Omega)/\Omega$ is plotted as a function of $\Omega$.
The value $\Omega$ for which $\pi \Gamma(\Omega)/\Omega=1$ (horizontal line) defines the Ioffe-Regel limit $\Omega^\text{IR}$.
}  
\label{tsparameters}
\end{figure*}
\begin{figure*}[]
\centering
\includegraphics[width=0.45\textwidth]{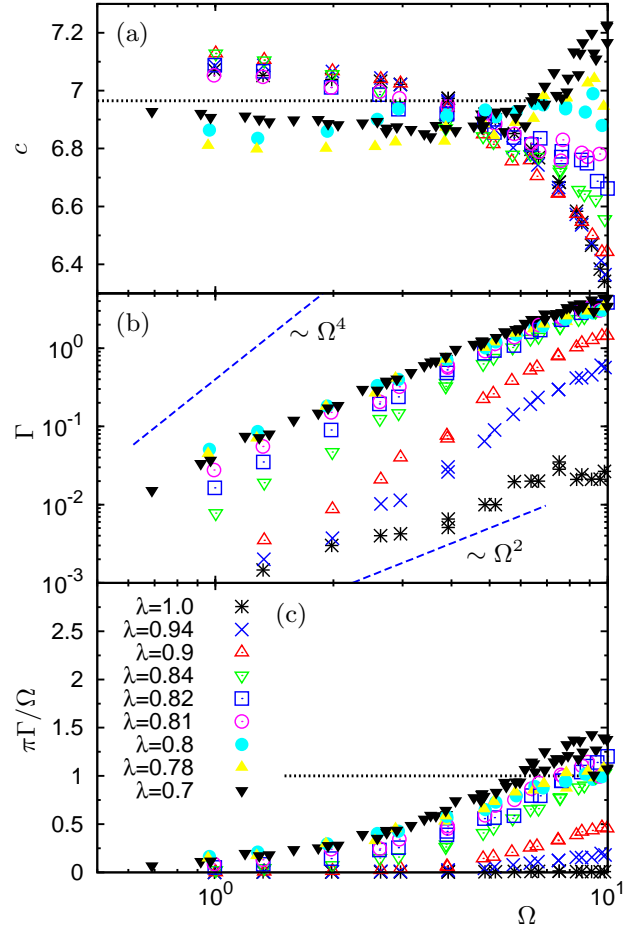}
\caption{
Spectroscopic parameters calculated from the longitudinal dynamic structure factors $S_{L}(\vec{q},\omega )$.
{\em a)} Phase velocity $c(\Omega)= \Omega(q)/q$, and {\em b)} broadening $\Gamma (\Omega)$, for the $L$ excitations in the direction $
[110]$, at the indicated values of $\lambda$. {\em c)} The ratio $\pi \Gamma(\Omega)/\Omega$ is plotted as a functions of $\Omega$ (see
 caption of Fig.~S\ref{tsparameters}).
} \label{lsparameters}
\end{figure*}
\end{document}